\def\etal{{\it et al.}}
\def\ie{{\it i.e.}}

\def\~{{$\tilde{\phantom{a}}$}}

\documentclass [12pt] {article}
\usepackage{epsfig}
\usepackage{color}

\def\thebibliography#1{\section{References}\markboth
 {REFERENCES}{REFERENCES}\list
 {[\arabic{enumi}]}{\settowidth\labelwidth{[#1]}\leftmargin\labelwidth
 \advance\leftmargin\labelsep
 \usecounter{enumi}}
 \def\newblock{\hskip .11em plus .33em minus -.07em}
 \sloppy
 \sfcode`\.=1000\relax}
\def\upcite#1{\raise6pt\hbox{\scriptsize
\cite{#1}}}
\def\lsim{\mathrel {\vcenter {\baselineskip 0pt \kern 0pt
    \hbox{$<$} \kern 0pt \hbox{$\sim$} }}}
\def\gsim{\mathrel {\vcenter {\baselineskip 0pt \kern 0pt
    \hbox{$>$} \kern 0pt \hbox{$\sim$} }}}
\def\gtlt{\mathrel {\vcenter {\baselineskip 0pt \kern 0pt
    \hbox{$>$} \kern 0pt \hbox{$<$} }}}

\def\hline{\noalign{\hrule \vskip2pt}}

\def\|{\ifmmode\Vert\else \char`\|\fi}
\ifx\oldzeta\undefined                          
  \let\oldzeta=\zeta                            
  \def\zzeta{{\raise 2pt\hbox{$\oldzeta$}}}     
  \let\zeta=\zzeta                              
\fi

\ifx\oldchi\undefined                           
  \let\oldchi=\chi                              
  \def\cchi{{\raise 2pt\hbox{$\oldchi$}}}       
  \let\chi=\cchi                                
\fi

\def\frac#1#2{{#1 \over #2}}

\def\half{\ifinner {\scriptstyle {1 \over 2}}
   \else {1 \over 2} \fi}

\def\simge{\mathrel{%
   \rlap{\raise 0.511ex \hbox{$>$}}{\lower 0.511ex \hbox{$\sim$}}}}
\def\simle{\mathrel{
   \rlap{\raise 0.511ex \hbox{$<$}}{\lower 0.511ex \hbox{$\sim$}}}}

\def\buildchar#1#2#3{{\null\!                   
   \mathop#1\limits^{#2}_{#3}                   
   \!\null}}                                    
\def\overcirc#1{\buildchar{#1}{\circ}{}}

\def\slashchar#1{\setbox0=\hbox{$#1$}           
   \dimen0=\wd0                                 
   \setbox1=\hbox{/} \dimen1=\wd1               
   \ifdim\dimen0>\dimen1                        
      \rlap{\hbox to \dimen0{\hfil/\hfil}}      
      #1                                        
   \else                                        
      \rlap{\hbox to \dimen1{\hfil$#1$\hfil}}   
      /                                         
   \fi}                                         %

\def\subrightarrow#1{
  \setbox0=\hbox{
    $\displaystyle\mathop{}
    \limits_{#1}$}
  \dimen0=\wd0
  \advance \dimen0 by .5em
  \mathrel{
    \mathop{\hbox to \dimen0{\rightarrowfill}}
       \limits_{#1}}}                           

\def\overlay#1#2{\ifmmode%
\setbox0=\hbox{$#1$}%
\setbox1=\hbox to\wd0{\hss$#2$\hss}\else%
\setbox0=\hbox{#1}%
\setbox1=\hbox to\wd0{\hss#2\hss}\fi%
#1\hskip-\wd0\box1 }

\def\pmb#1{\leavevmode\setbox0=\hbox{#1}%
\kern-.02em\copy0\kern-\wd0
\kern.04em\copy0\kern-\wd0
\kern-.02em\raise.04em\box0 }

\def\vereq#1#2{\lower3pt\vbox{\baselineskip1.5pt \lineskip1.5pt
\ialign{$\m@th#1\hfill##\hfil$\crcr#2\crcr\sim\crcr}}}

\def\tensor#1{\protect\@ontopof{#1}{\leftrightarrow}{1.15}\mathord{\box2}}
\def\overstar#1{\protect\@ontopof{#1}{\ast}{1.15}\mathord{\box2}}
\def\overdots#1{\protect\@ontopof{#1}{\cdots}{1.0}\mathord{\box2}}
\def\overcirc#1{\protect\@ontopof{#1}{\circ}{1.2}\mathord{\box2}}
\def\loarrow#1{\protect\@ontopof{#1}{\leftarrow}{1.15}\mathord{\box2}}
\def\roarrow#1{\protect\@ontopof{#1}{\rightarrow}{1.15}\mathord{\box2}}

\def\@ontopof#1#2#3{%
{\mathchoice
{\@@ontopof{#1}{#2}{#3}\displaystyle\scriptstyle}%
{\@@ontopof{#1}{#2}{#3}\textstyle\scriptstyle}%
{\@@ontopof{#1}{#2}{#3}\scriptstyle\scriptscriptstyle}%
{\@@ontopof{#1}{#2}{#3}\scriptscriptstyle\scriptscriptstyle}%
}%
}

\def\@@ontopof#1#2#3#4#5{%
\setbox0=\hbox{$#4#1$}%
\setbox1=\hbox{$#5#2$}%
\setbox2=\hbox{}\ht2=\ht0 \dp2=\dp0 %
\ifdim\wd0>\wd1 %
\setbox1=\hbox to\wd0{\hss\box1\hss}%
\mathord{\rlap{\raise#3\ht0\box1}\box0}%
\else   %
\setbox1=\hbox to.9\wd1{\hss\box1\hss}%
\setbox0=\hbox to\wd1{\hss$#4\relax#1$\hss}%
\mathord{\rlap{\copy0}\raise#3\ht0\box1}%
\fi
}%

\def\lambdabar{\protect\@lambdabar}
\def\@lambdabar{%
\relax
\bgroup
\def\@tempa{\hbox{\raise.73\ht0
\hbox to0pt{\kern.25\wd0\vrule width.5\wd0
height.1pt depth.1pt\hss}\box0}}%
\mathchoice{\setbox0\hbox{$\displaystyle\lambda$}\@tempa}%
{\setbox0\hbox{$\textstyle\lambda$}\@tempa}%
{\setbox0\hbox{$\scriptstyle\lambda$}\@tempa}%
{\setbox0\hbox{$\scriptscriptstyle\lambda$}\@tempa}%
\egroup
}

\def\corresponds{{\lower.2ex\hbox{=}}{\rm\kern-.75em^\triangle}}
\def\succsim{\succ\kern-.9em_\sim\kern.3em}
\def\precsim{\prec\kern-1em_\sim\kern.3em}
\def\slantfrac#1#2{\kern1em^{#1}\kern-.3em/\kern-.1em_{#2}}

\begin{document}
                                                                
\begin{center}
{\Large\bf Isotropic Radiators}
\\

\medskip

Haim Matzner
\\
{\sl Holon Academic Institute of Technology, Holon, Israel}
\\
Kirk T.~McDonald
\\
{\sl Joseph Henry Laboratories, Princeton University, Princeton, NJ 08544}
\\
(April 8, 2003)
\end{center}

\section{Introduction}

Can the radiation pattern of an antenna be isotropic?

A simple argument suggests that this is difficult.

The intensity of radiation depends on the square of the electric
(or magnetic) field.  To have isotropic radiation, it would seem
that the magnitude of the electric field would have to be
uniform over any sphere in the far zone.  However,
the electromagnetic fields in the far zone of an antenna are transverse,
and it is well known that a vector field of constant magnitude cannot
be everywhere tangent to the surface of a sphere (Brouwer's
``hairy-ball theorem'' \cite{Brouwer}).  Hence, it would appear
that the transverse electric field in the far zone cannot have the
same magnitude in all directions, and that the radiation pattern
cannot be isotropic, IF the radiation is everywhere linearly 
polarized \cite{Mathis,Saunders}.

However, electromagnetic waves can have two independent states of
polarization, described as elliptical polarization in the general
case.  While a transverse electric field with a single, linearly
polarized component cannot
be uniform over a sphere in the far zone, it may be possible that the
sum of the squares of the electric fields with two polarizations
is uniform.

\section{The U-Shaped Antenna of Shtrikman}

Shmuel Shtrikman has given an example of a U-shaped antenna
that generates an isotropic radiation pattern in the far zone
\cite{Matzner} in the limit of zero intensity of the radiation.
This example shows that any desired degree of ``isotropicity'' can be achieved
for a sufficiently weak radiation pattern.

Matzner \cite{Matzner2} has also shown that the radiation pattern of 
the U-shaped antenna can in principle be produced by specified
currents of finite strength on the surface of a sphere.

\subsection{The U-Shaped Antenna}

\begin{figure}[htp]  
\begin{center}
\vspace{0.1in}
\includegraphics[width=2.5in]{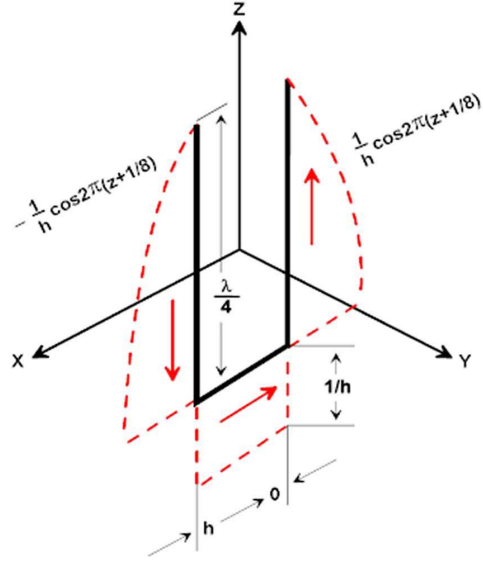}
\parbox{5.5in} 
{\caption[ Short caption for table of contents ]
{\label{fig1} The U-shaped antenna whose radiation pattern is isotropic
in the limit that $h \to 0$, for which the intensity also goes to zero.
The dashed lines indicate the spatial dependence of the current.
From \cite{Matzner}.
}}
\end{center}
\end{figure}

The U-shaped antenna of Matzner \etal\ \cite{Matzner}
is illustrated in Fig.~\ref{fig1}.
It consists of two vertical arms of length $L = \lambda / 4$ $(k L = \pi / 2)$,
separated by a short cross piece of length $h \ll \lambda$.

Denoting the peak current in the antenna by $I$, the current density {\bf J}
can be written
\begin{equation}
{\bf J}({\bf r},t) = {\bf J}({\bf r}) e^{- i \omega t},
\label{eq2}
\end{equation}
where
\begin{eqnarray}
{\bf J}({\bf r}) & = & I \left[ 
\delta(x - h / 2) \delta(y) \cos(k z + \pi / 4) \hat{\bf z} \right.
+\ \delta(y) \delta(z + \lambda/8) \hat{\bf x} 
\nonumber \\
& & \left.
-\ \delta(x + h / 2) \delta(y) \cos(k z + \pi / 4) \hat{\bf z}
\right],
\label{eq3}
\end{eqnarray}
and $-\lambda/8 \leq z \leq \lambda/8$ on the vertical arms, $-h/2 \leq x \leq h/2$
on the horizontal arm.

The time-averaged, far-zone radiation pattern of an antenna with a specified, 
time-harmonic current 
density can be calculated (in Gaussian units) according to \cite{Panofsky}
\begin{equation}
{d P \over d \Omega} = {\omega^2 \over 8 \pi c^3} \left| \hat{\bf k} \times \left[ 
\hat{\bf k} \times \int {\bf J}({\bf r}) e^{i {\bf k} \cdot {\bf r}} d{\rm Vol}
\right] \right|^2.
\label{eq4}
\end{equation}

For an observer at angles $(\theta,\phi)$ with respect to the $z$ axis (in a
spherical coordinate system), the unit wave vector has rectangular components
\begin{equation}
\hat{\bf k} = \sin\theta \cos\phi\ \hat{\bf x} + \sin\theta \sin\phi\ \hat{\bf y}
+ \cos\theta\ \hat{\bf z}.
\label{eq5}
\end{equation}
The integral transform ${\bf J}_k = \int {\bf J}({\bf r}) e^{i {\bf k} \cdot 
{\bf r}} d{\rm Vol}$
in eq.~(\ref{eq4}) has rectangular components
\begin{eqnarray}
J_{k,x} & = &
I \int_{-h/2}^{h/2} dx e^{i k \sin\theta\cos\phi x}
\int dy \delta(y) e^{ i k \sin\theta\sin\phi y}
\int dz \delta(z + 1/8) e^{i k \cos\theta z}
\nonumber \\
& = & I {\sin [(k/2) h \sin\theta \cos\phi] \over (k/2) \sin\theta\cos\phi} 
e^{-i (\pi / 4) \cos\theta}
\approx I h e^{-i (\pi / 4) \cos\theta},
\label{eq6} \\
J_{k,y} & = & 0,
\label{eq7} \\
J_{k,z} & = &
I \int dx \delta(x - h/2) e^{i k \sin\theta\cos\phi x}
\int dy \delta(y) e^{i k \sin\theta\sin\phi y}
\int_{-\lambda/8}^{\lambda/8} dz \cos(k z + \pi / 4) e^{i k \cos\theta z}
\nonumber \\
& & -I \int dx \delta(x + h/2) e^{i k \sin\theta\cos\phi x}
\int dy \delta(y) e^{i k \sin\theta\sin\phi y}
\int_{-\lambda/8}^{\lambda/8} dz \cos(k z + \pi / 4) e^{i k \cos\theta z}
\nonumber \\
& = & I {\sin[(k/2) h \sin\theta \cos\phi] \over (k/2) \sin^2\theta}
(i e^{i (\pi/4) \cos\theta} + \cos\theta e^{-i (\pi/4) \cos\theta})  
\nonumber \\
& \approx & I h { \cos\phi \over \sin\theta}
(i e^{i (\pi/4) \cos\theta} + \cos\theta e^{-i (\pi/4) \cos\theta}). 
\label{eq8}
\end{eqnarray}
Then,
\begin{eqnarray}
\left| \hat{\bf k} \times \left( \hat{\bf k} \times {\bf J}_k \right) \right|^2 & = &
| {\bf J}_k |^2 - | \hat{\bf k} \cdot {\bf J}_k |^2
\nonumber \\
& = & (1 - \hat k_x^2) |J_{k,x}|^2  +  (1 - \hat k_z^2) |J_{k,z}|^2 
- 2 \hat k_x \hat k_z Re J_{k,x} J^\star_{k,z} 
\nonumber \\
& = &  (1 - \sin^2\theta \cos^2\phi) | J_{k,x} |^2  +  \sin^2\theta | J_{k,z} |^2 
- 2 \sin\theta \cos\theta \cos\phi Re J_{k,x} J^\star_{k,z}
\nonumber \\
& = & I^2 {\sin^2 [(k/2) h \sin\theta \cos\phi] \over [(k/2) \pi \sin\theta\cos\phi]^2}
\left[ (1 - \sin^2\theta \cos^2\phi) \right.
\nonumber \\
& & \qquad + \cos^2\phi (1 + \cos^2\theta - 2 \cos\theta \sin [(\pi / 2) \cos\theta] )
\nonumber \\
& & \qquad \left. - 2 \cos^2\phi (\cos^2\theta - \cos\theta \sin [(\pi / 2) \cos\theta])
\right]
\nonumber \\
& = & I^2 {\sin^2 [(k/2) h \sin\theta \cos\phi] \over [(k/2) \sin\theta\cos\phi]^2}
\approx I^2 h^2.
\label{eq9}
\end{eqnarray}
Thus, the radiation pattern is indeed isotropic in the limit that $h \to 0$.
But in this limit, the radiation vanishes, for a fixed peak current $I$.\footnote{Matzner
\etal\ \cite{Matzner} tacitly assume that the product $Ih = 1$ as $h \to 0$.  Their result
then appears to have a finite radiation intensity, but the current in their 
U-shaped antenna is infinite.}

For a finite separation $h$ between the two vertical arms of the antenna, the deviation 
from isotropicity is roughly $1 - \sin^2 (k h/2) / (k h / 2)^2$. 
Thus the pattern will be isotropic to 1\% for $h \sim 0.05 \lambda$.  However, this
uniformity is achieved at the expense of a substantial reduction in the intensity of the radiation.
For example, the case of a U-shaped antenna with $h = 0.05 \lambda$ has an intensity only
1/40 of that of a basic half-wave, center-fed antenna.\footnote{Using eq.~(14-55) of
\cite{Panofsky} for a center-fed linear antenna of length $L = \lambda / 2$
($kL = \pi$), and peak current $I$, we have
\begin{equation}
\left| \hat{\bf k} \times \left( \hat{\bf k} \times {\bf J}_k \right) \right|^2 =
{4 I^2 \over k^2} \left[ {\cos[(kL/2) \cos\theta] - \cos(kL/2) \over \sin\theta \sin(kL/2)}
\right]^2
= {I^2 \over \pi^2} {\cos^2[(\pi/2) \cos\theta] \over \sin^2\theta }\, .
\label{eq10}
\end{equation}
for which the maximum intensity occurs at $\theta = \pi / 2$ where 
eq.~(\ref{eq10}) becomes $0.10 I^2$.}

\begin{figure}[htp]  
\begin{center}
\vspace{0.1in}
\includegraphics[width=4.5in]{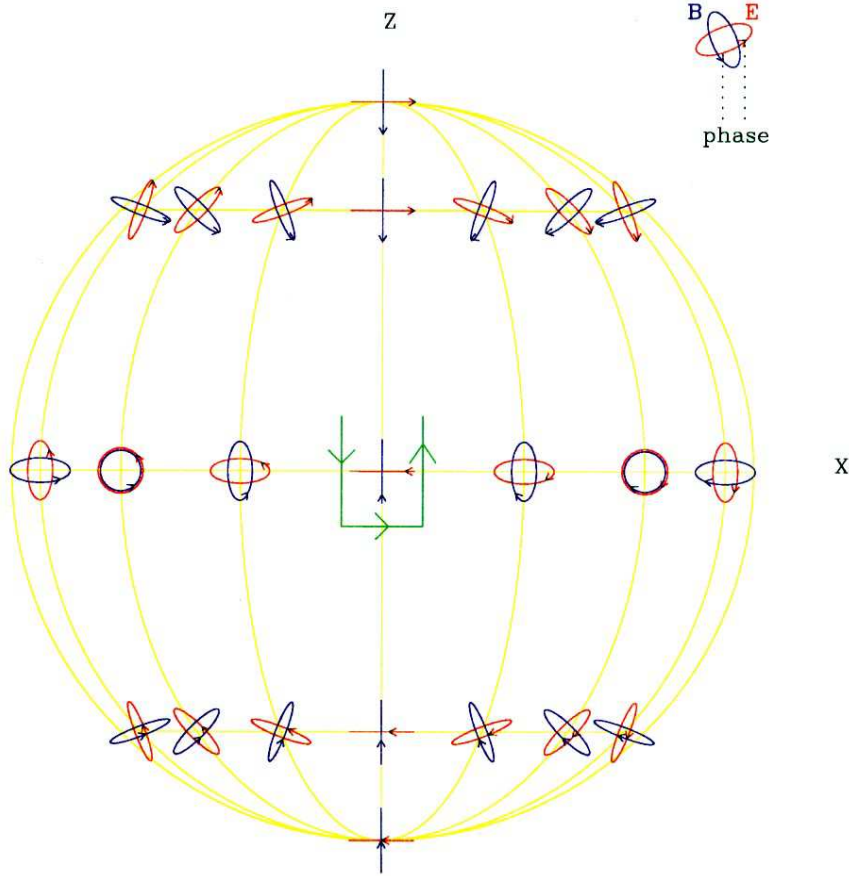}
\parbox{5.5in} 
{\caption[ Short caption for table of contents ]
{\label{fig2} The elliptical polarization of the fields
(\ref{eq13})-(\ref{eq14}) of the U-shaped antenna
in the limit of small $h$.  From \cite{Matzner2}.
}}
\end{center}
\end{figure}

As is to be expected, the polarization of the radiation of the 
U-shaped antenna is elliptical in general.  The far-zone electromagnetic
fields are related to the integral transform ${\bf J}_k$ according to
\begin{equation}
{\bf B} = i k {e^{i (kr - \omega t)} \over r} \hat{\bf k} \times {\bf J}_k,
\qquad
{\bf E} = {\bf B} \times \hat{\bf k}.
\label{eq11}
\end{equation}
The components of the far-zone electromagnetic fields in spherical coordinates are
therefore,
\begin{eqnarray}
E_r = B_r & = & \hat{\bf k} \cdot {\bf B} = 0,
\label{eq12} \\
E_\theta = B_\phi & = & i k {e^{i (kr - \omega t)} \over r}
(\cos\theta \cos\phi J_{k,x} - \sin\theta J_{k,z})
\nonumber \\
& = & I k \cos\phi  e^{i (\pi / 4) \cos\theta} 
 {\sin [(k/2) h \sin\theta \cos\phi] \over (k/2) \sin\theta\cos\phi} 
{e^{i (kr - \omega t)} \over r}
\nonumber \\
& \approx &  I h k \cos\phi  e^{i (\pi / 4) \cos\theta} 
{e^{i (kr - \omega t)} \over r}\, ,
\label{eq13} \\
E_\phi = - B_\theta & = & - i k {e^{i (kr - \omega t)} \over r} \sin\phi J_{k,x}
\nonumber \\
& = & - i I k \sin\phi  e^{-i (\pi / 4) \cos\theta} 
 {\sin [(k/2) h \sin\theta \cos\phi] \over (k/2) \sin\theta\cos\phi} 
{e^{i (kr - \omega t)} \over r}
\nonumber \\
& \approx & - i I h k \sin\phi  e^{-i (\pi / 4) \cos\theta} 
{e^{i (kr - \omega t)} \over r}\, .
\label{eq14}
\end{eqnarray}
The magnitudes of the fields are
\begin{equation}
E = B = {I k \over r} 
{\sin [(k/2) h \sin\theta \cos\phi] \over (k/2) \sin\theta\cos\phi}
\approx {I h k \over r}\, ,
\label{eq15}
\end{equation}
which are isotropic in the limit of small $h$.  Figure~\ref{fig2}
from \cite{Matzner2} illustrates the character of the
elliptical polarization of the fields (\ref{eq13})-(\ref{eq14})
for various directions in the limit of small $h$.

\subsection{Isotropic Radiation from Currents on a Spherical Shell}

In sec.~6.6 of his Ph.D.\ thesis \cite{Matzner2}, 
Matzner shows how the far-zone radiation pattern of the
U-shaped antenna (in the limit $h \to 0$) can be reproduced by an
appropriate distribution of currents on a spherical shell of radius 
$R = \lambda /4$.  For this, he first expands the far-zone fields
(\ref{eq13})-(\ref{eq14}) in vector spherical harmonics, and
then matches these to currents on a shell of radius $R$ and
to an appropriate form for the fields inside the shell.

\begin{figure}[htp]  
\begin{center}
\vspace{0.1in}
\includegraphics[width=2.5in]{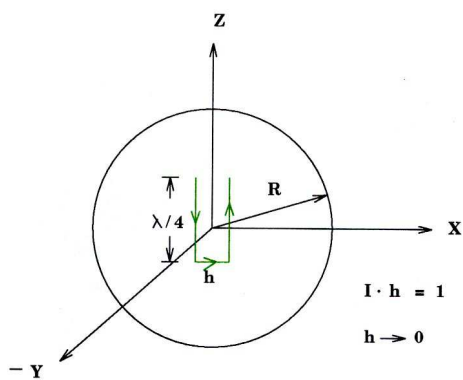}
\parbox{5.5in} 
{\caption[ Short caption for table of contents ]
{\label{fig3} The spherical shell of radius $R = \lambda / 4$
on which a set of currents can be found that produces the
same far-zone fields as does the U-shaped antenna.
From \cite{Matzner2}.
}}
\end{center}
\end{figure}

\begin{figure}[htp]  
\begin{center}
\vspace{0.1in}
\includegraphics[width=5in]{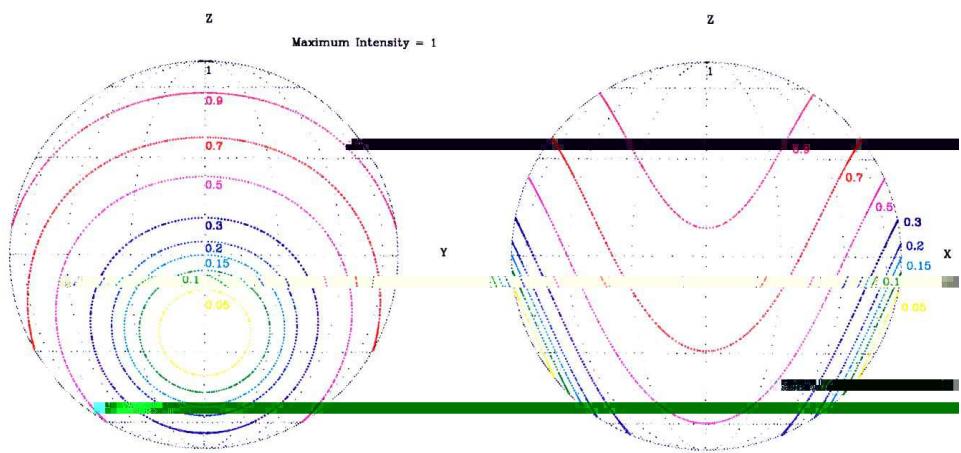}
\parbox{5.5in} 
{\caption[ Short caption for table of contents ]
{\label{fig4} Contours of the current density on the spherical
shell that produce the
same far-zone fields as does the U-shaped antenna.
From \cite{Matzner2}.
}}
\end{center}
\end{figure}

Figures \ref{fig3} and \ref{fig4} illustrate this procedure.
The key point is that the surface currents are finite in 
magnitude, and hence an isotropic radiator is realizable 
in the laboratory (in contrast to the U-shaped antenna, 
which requires an infinite current $I$ to achieve perfectly
isotropic radiation).

In principle, many other surfaces besides that of a sphere could
support a pattern of finite, oscillating currents whose far zone
radiation pattern is isotropic.


\section{A Linear Array of ``Turnstile'' Antennas}

Saunders \cite{Saunders} has noted that a certain infinite array
(a certain vertical stack) 
of so-called ``turnstile'' antennas \cite{Brown,Cebik} can also
produce a far-zone radiation pattern that is isotropic

\begin{figure}[htp]  
\begin{center}
\vspace{0.1in}
\includegraphics[width=3in]{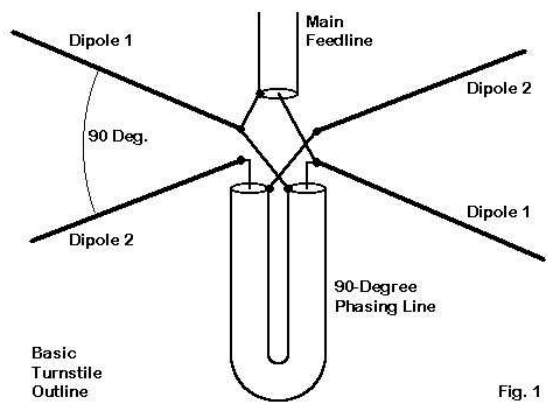}
\parbox{5.5in} 
{\caption[ Short caption for table of contents ]
{\label{fig5} A ``turnstile'' antenna.
From \cite{Cebik}.
}}
\end{center}
\end{figure}

A turnstile antenna
consists of a pair of half-wave, center-fed linear dipole
antennas oriented at 90$^\circ$ to each other, and driven
90$^\circ$ out of phase, as shown in Fig.~\ref{fig5}.

If we approximate the half-wave dipoles by point dipoles, then
the dipole moment of the system can be written
\begin{equation}
{\bf p} = {\bf p}_0 e^{- i \omega t}
= p_0 (\hat{\bf x} + i \hat{\bf y}) e^{- i \omega t},
\label{eq16}
\end{equation}
taking the antenna to be aligned along the $x$ and $y$ axes.
The electromagnetic fields in the far zone are then
\begin{equation}
{\bf B} =  k^2 {e^{i (kr - \omega t)} \over r} \hat{\bf k} \times {\bf p}_0,
\qquad
{\bf E} = {\bf B} \times \hat{\bf k},
\label{eq17}
\end{equation}
whose components in spherical coordinates are
\begin{eqnarray}
E_r = B_r & = & \hat{\bf k} \cdot {\bf B} = 0,
\label{eq18} \\
E_\theta = B_\phi & = & - p_0 k^2 {e^{i (kr - \omega t)} \over r}
\cos\theta (\cos\phi + i \sin\phi),
\label{eq19} \\
E_\phi = - B_\theta & = & - p_0 k^2 {e^{i (kr - \omega t)} \over r} 
(\sin\phi - i \cos\phi).
\label{eq20}
\end{eqnarray}
In the plane of the antenna, $\theta = 90^\circ$, the electric
field has no $\theta$ component, and hence no $z$ component; the turnstile 
radiation in the horizontal plane is horizontally polarized.
In the vertical direction, $\theta = 0^\circ$ or $180^\circ$, the
radiation is circularly polarized.  For intermediate angles $\theta$
the radiation is elliptically polarized.

\begin{figure}[htp]  
\begin{center}
\vspace{0.1in}
\includegraphics[width=4in]{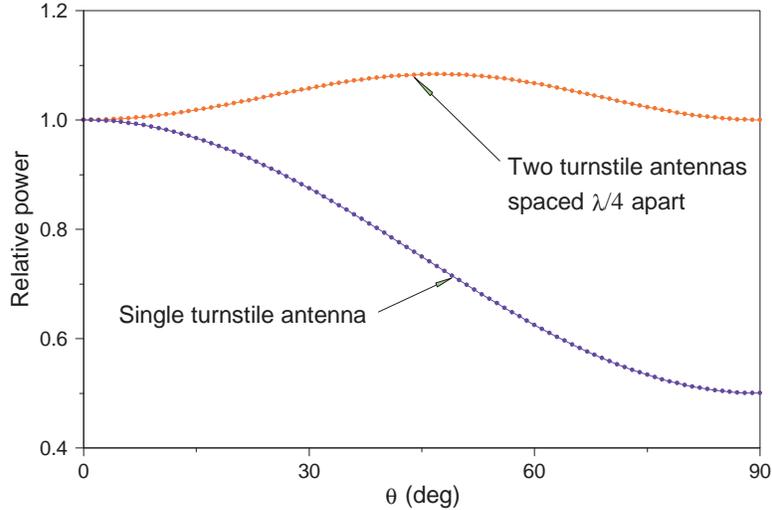}
\parbox{5.5in} 
{\caption[ Short caption for table of contents ]
{\label{fig6} The relative radiation pattern of a single turnstile
antenna, and of a pair of turnstile antennas that are separated
by $\lambda/4$.
}}
\end{center}
\end{figure}

The magnitudes of the fields are
\begin{equation}
E = B = {p_0 k^2 \over r} \sqrt{1 + \cos^2\theta},
\label{eq21}
\end{equation}
so the time-averaged radiation pattern is
\begin{equation}
{d P \over d \Omega} = {c r^2 \over 8 \pi} B^2
= {p_0^2 \omega^4 \over 8 \pi c^3} (1 + \cos^2\theta).
\label{eq22}
\end{equation}
The intensity of the radiation varies by a factor of 2
over the sphere, that is, by 3 dB, as shown in Fig.~\ref{fig6}.
Compared to other
simple antennas, this pattern is remarkably isotropic.

But we can make the pattern even more isotropic by considering a
vertical stack of turnstile antennas.

If the center of the turnstile antenna had been at height $z$
along the $z$-axis, the only difference in the resulting
electric and magnetic fields would be a phase change by
$k z \cos\theta$ because the path length to the distant
observer differs by $z \cos\theta$.  That is, the fields
(\ref{eq18})-(\ref{eq20}) would simply be multiplied by
the phase factor $e^{- i k z \cos\theta}$.

Thus, if we have two turnstile antennas, one whose center is at the
origin, and the other whose center is at height $z$, and we operated
them in phase, the fields (\ref{eq18})-(\ref{eq20}) would be 
multiplied by   
\begin{equation}
1 + e^{- i k z \cos\theta}.
\label{eq23}
\end{equation}
The radiated power would
therefore by eq.~(\ref{eq22}) multiplied by the absolute square of
eq.~(\ref{eq23}):
\begin{equation}
{d P \over d \Omega} = 2 {p_0^2 \omega^4 \over 8 \pi c^3} (1 + \cos^2\theta)
[ 1 + \cos(k z \cos\theta)].
\label{eq24}
\end{equation}
For example, suppose $kz = \pi / 2$, \ie, the vertical separation of the
two antennas is 1/4 of a wavelength.  Then, the peak of the radiation
pattern is only 1.08 times (0.35 db) greater than the minimum, as shown in
Fig.~\ref{fig6}.  For most practical purposes, this double turnstile
antenna could be considered to be isotropic.

\begin{figure}[htp]  
\begin{center}
\vspace{0.1in}
\includegraphics[width=3in]{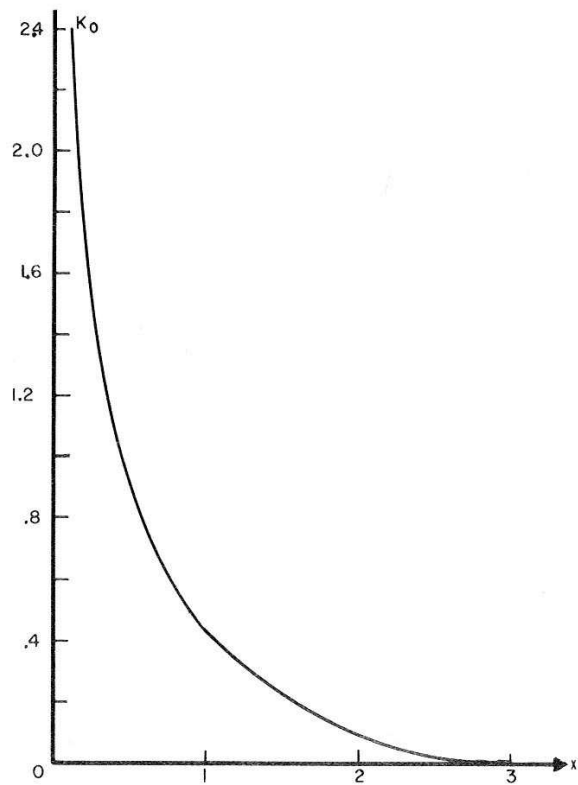}
\parbox{5.5in} 
{\caption[ Short caption for table of contents ]
{\label{fig7} The modified Bessel function $K_0(x)$.
From \cite{Abramowitz}.
}}
\end{center}
\end{figure}

Saunders \cite{Saunders}
has further shown that an infinite array of turnstile antennas
yields strictly isotropic radiation provided the number $N(z)$ of such 
antennas in an interval $dz$ along the vertical axis is proportional to
$K_0(kz)$, the so-called modified Bessel function of order zero
\cite{Abramowitz}, whose behavior is sketched in Fig.~\ref{fig7}.
The antennas are all driven in phase.  Since the function $K_0(kz)$ is
sharply peaked at $z = 0$, we see that a properly spaced
collection of turnstile
antennas that extends over only $\pm 1$ wavelength in $z$ could
produce an extremely isotropic radiation pattern.

The U-shaped antenna of sec.~2 is a variant on the theme of a vertical stack of 
turnstile antennas.  Since the currents are opposite in the two vertical arms 
of the U-shaped antenna, the charge accumulations on these arms have
opposite signs as well.  Thus, the two vertical arms are in effect a vertical
stack of horizontal dipole antennas.  If we had a second U-shaped
antenna, rotated by 90$^\circ$ about the vertical compared to the first,
and driven $90^\circ$ out of phase, this would be equivalent to a vertical
stack of (horizontal) turnstile antennas.  Such a double U-shaped antenna is discussed
in sec.~6.5.2 of \cite{Matzner2}, where its radiation pattern is found
to be isotropic, although the details of the polarization of the radiation
fields differ slightly from those for a single U-shaped radiator.



\begin{thebibliography}{99}

\bibitem{Brouwer}
L.E.J.~Brouwer, 
{\sl On Continuous Vector Distributions on Surfaces},
Proc.\ Royal Acad.\ (Amsterdam) {\bf 11}, 850 (1909);
{\em Collected Works},
{\sl Volume 2: Geometry, Analysis, Topology, and Mechanics},
ed.\ by Hans Freudenthal
(North-Holland Publishing Company, 1976), p.~301.

\bibitem{Mathis}
H.F.~Mathis,
{\sl A short proof that an isotropic antenna is impossible},
Proc.\ I.R.E.\ {\bf 39}, 970 (1951);
{\sl On isotropic antennas},
Proc.\ I.R.E.\ {\bf 42}, 1810 (1954).

\bibitem{Saunders}
W.K.~Saunders,
{\sl On the Unity Gain Antenna},
in {\em Electromagnetic Theory and Antennas},
ed.\ by E.C.~Jordan
(Pergamon Press, New York, 1963), Vol.~2, p.~1125.

\bibitem{Matzner}
H.~Matzner, M.~Milgrom and S.~Shtrikman,
{\sl Magnetoelectric Symmetry and Electromagnetic Radiation},
Ferroelectrics {\bf 161}, 213 (1994).

\bibitem{Matzner2}
H.~Matzner,
{\em Moment Method and Microstrip Antennas},
Ph.D.\ Thesis (Weizmann Institute of Science, Rehovot, Israel, 1993).

\bibitem{Panofsky}
See, for example, eq.~(14-53) of
W.K.H.~Panofsky and M.~Phillips,
{\sl Classical Electricity and Magnetism},
2nd ed.\ (Addison-Wesley, Reading, MA, 1962).

\bibitem{Brown}
G.H.~Brown,
{\sl The ``Turnstile'' Antenna},
Electronics {\bf 9}, 15 (April, 1936).

\bibitem{Cebik}
L.B.~Cebik,
{\sl The Turnstile Antenna. An Omni-Directional
Horizontally Polarized Antenna}, 
http://www.cebik.com/turns.html

\bibitem{Abramowitz}
M.~Abramowitz and I.~Stegun,
{\em Handbook of Mathematical Functions}
(National Bureau of Standards, Washington, D.C., 1964), sec.~9.6.

\end{thebibliography}
\end{document}